# Text Based Approach For Indexing And Retrieval Of Image And Video: A Review

Avinash N Bhute  and  B.B. Meshram

VJTI, Matunga, Mumbai-19

## ABSTRACT

*Text data present in multimedia contain useful information for automatic annotation, indexing. Extracted information used for recognition of the overlay or scene text from a given video or image. The Extracted text can be used for retrieving the videos and images. In this paper, firstly, we are discussed the different techniques for text extraction from images and videos. Secondly, we are reviewed the techniques for indexing and retrieval of image and videos by using extracted text.*

## 1. INTRODUCTION

Text data present in multimedia viz. video and images contain useful information for automatic annotation, indexing. The Process of Extraction of information is detection, localization, tracking, extraction, enhancement, and recognition of the text from a given image [1]. However, there are differences in text in style, orientation, size, and alignment, as well as low contrast image and complex background make the automatic text extraction problem more difficult and time consuming. While critical surveys of related problems such as document analysis face detection and image & video indexing and retrieval can be found, the problem of text extraction isn't surveyed well. A variety of approaches to text extraction from images and video have been presented for many applications like address block location [6], content-based image/video indexing [2,8], page segmentation [4,5],  and license plate location [3,7]. In spite of such in critical studies, it is still not easy to design a general-purpose Text Extraction system. This is often a result of so many possible sources of variation once extracting text from complex images, or from images having difference in style, color, orientation, font size and alignment.

Although images non-inheritable by scanning book covers, CD covers, or different multi-colored documents have almost similar characteristics as the document images, Author(s)  cannot be dealt with conventional document image analysis technique. Text in video images can classify into caption text and scene text. The caption text is artificially overlaid on the image and scene text exists naturally in the images. Some researchers prefer to use the term 'graphics text' for scene text, and 'superimposed text' or 'artificial text' for caption text [9,10]. It is documented that scene text is harder to detect.

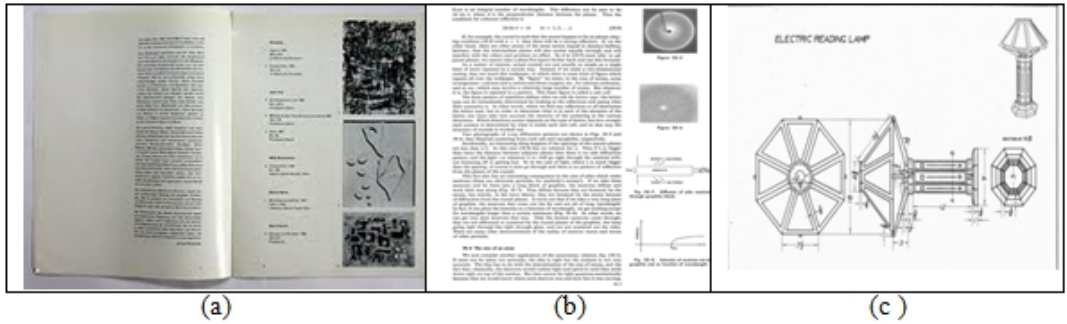

Fig. 1 Grayscale document images: (a) single-column text from a book, (b) two-column page from a book
(c) Electric Lamp Drawing

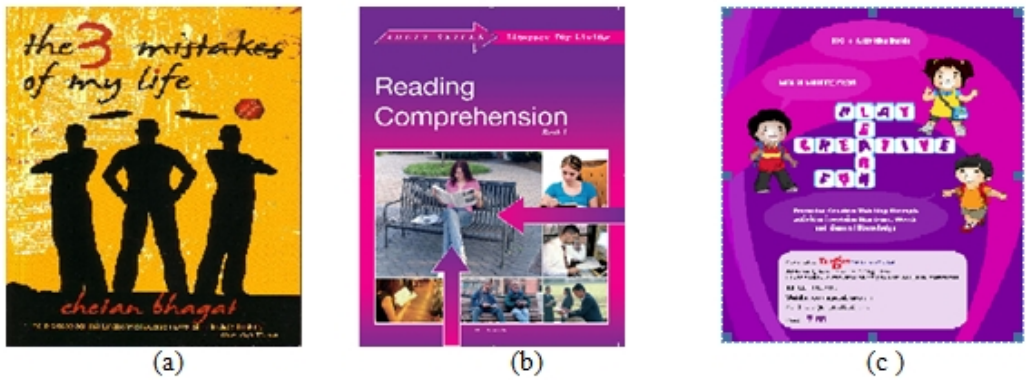

Fig. 2. Multi-color document images: each text line may or may not be of the same color.

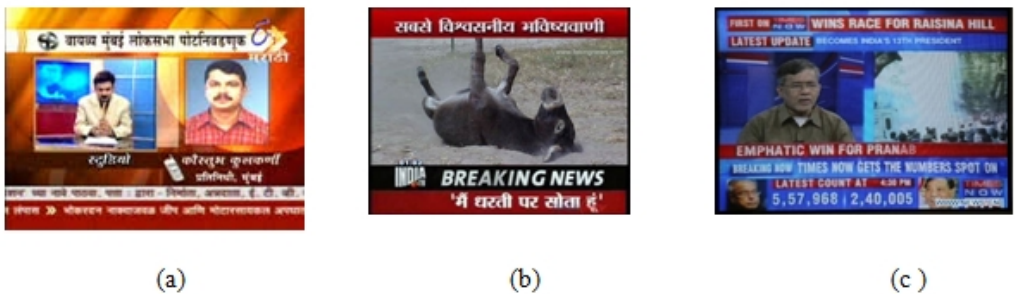

Fig. 3 Images with Caption Text

A Text Extraction system accepts an input as a still image or a sequence of frames. The frames can be in gray scale or color, compressed or un-compressed, and also the text within the frames could or might not move. Text in frames will exhibit many variations according to their properties such as Geometry (size, alignment, inter-character distance), Color( monochrome, polychrome), Motion(static, linear moment), Edge(text boundaries, strong edges), Compression, etc.

Organization of paper In section-I we are introduced the concept of text extraction from images/video as well as text based image video retrieval. Section two discussed the text extraction from images and videos with its process such as Text Tracking (TT), Text Detection (TD), Text Localization (TL), Text Extraction and Enhancement (TEE), Optical Character Recognition (OCR). Section-III deals with techniques and approaches for text based image and video indexing and retrieval systems. Finally Section IV has summary followed by conclusion.

## 2. TEXT EXTRACTION FROM IMAGES AND VIDEOS

The process for extraction carries detection, localization, tracking, extraction, enhancement, and recognition of the text [1] from a given image as shown in figure 1. Text detection refers to the identification of text in a given video frame. Text localization is refer to determine location of text in the frame or sequence of video frame. Text tracking is performed to cut back the interval for text localization and to maintain the position across adjacent frames. The precise location of text in a frame indicated by bounding boxes, the text still has to be divided from the background to useful for its recognition. Text extraction is that the stage wherever the text components are segmented from the background. Extracted text components are required enhancement because the text region usually has low-resolution and is susceptible to noise. The extracted text images are transformed into plain text using OCR technology.

### 2.1 Text Detection

The text of input images need to be identified as the input image contains any text, the existence or non-existence of text among the image. Several approaches assuming certain types of video frame or image contain text (e.g., recording cases or book covers). However, in the case of video, the amount of frames containing text is far smaller than the amount of frames while not text. The text detection stage detects the text in image. Kim [11] chosen a frame from shots detected by a scene-change detection methodology as a candidate containing text. Kim's scene-change detection methodology isn't delineated thoroughly in his literature, he has mention that low threshold values are required for scene-change detection as a result of a very small region of images the occupied by text regions. This approach is incredibly sensitive to scene-change detection. Text-frame choice is performed at associate interval of two seconds for caption text within the detected scene frames. This can be an easy and economical resolution for video indexing applications that only needs keywords from video clips, instead of the whole text.

Smith and Kanade [12] detected the text information based on scene-change which allows the distinction between two consecutive. Author(s) achieved accuracy of 90% in scene-change detection. Gargi et al. [13,14] performed text detection by assuming the amount of intra-coded blocks in P- and B- frames of an MPEG compressed video will increase, once a text caption appears. Lim et al. [15] made an easy assumption that text typically includes a higher intensity than the background. Author(s) counted the number pixels that are less than a threshold and exhibited a big color distinction to their neighborhood. This methodology is very simple and fast. However, issues will occur with color-reversed text. A text localization used for detecting the text in image/video.

Zhong et al. [16], Antani et al. [17] presented text localization on compressed images, that resulted in an exceedingly quicker performance. Therefore, their text localizers might even be used for detection of text. The process of text detection stage is related with the text localization and text tracking stage. Google Image labeler [63] is It is feature in the form of game which help to improve the quality of Google's image search results and allowed the user to label to images.

It was online from 2006 to 2011 design simply for fun. It was way for Google to ensure that its keywords were matched to correct images. Each matched word help Google to build an accurate database for Google Image Search.

## 2.2 Text Localization

Text localization techniques can be further divided into two types based on features utilization: region-based and texture-based. Region-based technique use the properties of the color/gray scale in a text region or their variations with the corresponding properties of the background. These technique can be additional divided into 2 sub-approaches: connected component(CC) and edge-based. These two approaches works in a bottom-up fashion; by distinctive sub-structures, like CCs or edges, and so integrating these structures to mark rectangles for text. CC-based methods are groups the small parts into larger parts till all regions are identified in the image. All text parts using spatial arrangement require geometrical analysis and on separate non-text parts mark the limits of the text regions.

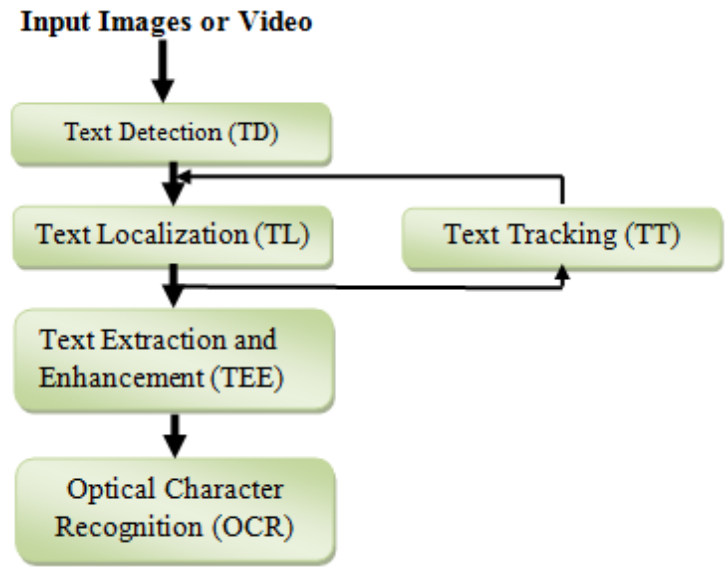

Fig. 4. Text Extraction from images/videos Architecture

Ohya et al. [18] presented a four-stage process: (i) binarization based on local thresholding, (ii) character component detection using gray-level difference, (iii) character recognition and (iv) relaxation operation to update the similarities. Author(s) were able to extract and recognize characters, together with multi-segment characters, under varied illuminating conditions, sizes, positions, and fonts once handling scene text images, like signboard, train, advt., etc.

However, binary segmentation is no appropriate for video documents, which might have many objects with completely different grey levels and high levels of noise and differences in illumination. Moreover, this approach places many restrictions associated with text alignment, like upright and not connected, also because the color of the text (monochrome). Lee and Kankanhalli [19] presented CC-based methodology to detect and recognize of text on cargo containers, Which has uneven lighting conditions and completely different sizes and shapes characters. Edge information is employed for a coarse search before the CC generation. The boundaries of potential characters are determined by the difference between adjacent pixels an input image. Threshold values are chosen for every text candidate based on pixels on the boundaries. Despite their claims that the method may be effectively utilized in different domains, experimental results were solely conferred for cargo container images.

Zhong et al. [20] prepared a CC-based methodology, that uses color reduction. Author(s) quantize the color using the peaks in a color histogram (RGB Color Space). It is Assume that the text regions cluster togather during this color space and occupy good portion of a image. Every text component goes through a filtering stage employing a variety of heuristics, like spatial alignment ,space and diameter. The performance of this technique was evaluated using CD images and book cover images.

Kim [11] presented the approach by segments an image in the RGB space using color clustering in a color histogram. Non-text components. The horizontal text lines associated text segments are extracted based on iterative projection profile analysis. After that these text segments are integratedby heuristics. Since many threshold values ought to be determined by trial and error, this approach isn't appropriate as a general-purpose text localizer. Experiments were performed with fifty video images, together with numerous character sizes and designs, and a localization rate of eighty seven was reported. Lienhart et al. [21,22] regard text regions as CCs with similar size and color, and apply motion analysis to boost the text extraction results. Segments that are too small and overlarge are filtered out. After dilation, For enhance and extracted results motion information and contrast analysis are used. Their main focus is on caption text, that exhibit the higher contrast with the background. Finally, a geometrical analysis, together with the dimension, height, and ratio, is employed to separate any non-text parts.

Shim et al. [8] proposed the homogeneity of intensity of text regions in images. Jain and Yu [23] apply a CC-based methodology once preprocessing, which incorporates bit dropping, color clump, ambiguous image decomposition, and foreground image generation. Once the input frame is decomposed into multiple foreground frames, every foreground image goes through identical text localization stage. By using block adjacency graph, CCs are generated in parallel for all the foreground images. The localized text component within the individual foreground video frame is then integrated into one output frame. The algorithm extracts solely horizontal and vertical text, and not skew text. The authors imply that their algorithm might not work well once the color histogram is sparse.

Messelodi and Modena's technique [24] consists of 3 serial stages: (i) extraction of elementary objects, (ii) filtering of objects, and (iii) text line selections. Preprocessing, like noise reduction, de-blurring, contrast enhancement, quantization, etc., is additionally performed. After the preprocessing, intensity normalization, image binarization, and CC generation are performed. The authors notices that the filter selection is heavily depends on and threshold values. Finally, the text line selection stage starts from one region and recursively expands, till a termination criterion is satisfied. Kim et al. [25] used cluster-based templates for sort out non-character components. The similar approach was additionally reported by Ohya et al. [34].

Hase et al. [26] proposed a CC-based technique for color documents. Author(s) assume that each character is printed in a single color. The image is then divided into many binary images and string extraction by multi-stage relaxation that was given by authors previously [27] is performed on every binary image. The character strings are selected by their likelihoods, using conflict resolution rules by merging all results from the individual binary images, The inclusion and overlap, the conflicts between character strings are filtered using tree illustration. Contrary to other text localization techniques, Author(s) pay additional attention to the filtering stage. As a result, Author(s) deals with shadowed and curved strings

CC-based methods are widely used uue to their relatively simple implementation. Almost all CC-based methods have four processing stages: (i) preprocessing, like color clustering and noise reduction, (ii) CC generation, (iii) filtering out non-text components, and (iv) Component grouping [11]. Further, the performance of a CC-based technique is affected by component grouping, like a projection profile analysis or text line selection. Additionally, many threshold values are required to filter out the non-text components, and these threshold values are depends on the image/video database.

Binarization techniques, that use global, local, or adaptive thresholding, are the simplest methods for text localization. This approach has been adopted for specific applications like address location on snail mail, courtesy amount on checks, etc., thanks to its simplicity in implementation [28].

Due to the variety of possible variations in text in different types of applications and also the limitations of any single approach to wear down such variations. Zhong et al. [20] presented integration of CC-based approach and the texture-based approach. If characters are merged or not well separated from the background, CC-based technique doesn't perform well. Additionally the disadvantage of their texture-based technique is that characters, that extend below the baseline or above the characters, are segmented into 2 Component. In the hybrid scheme the CC-based technique is applied once the bounding boxes are localized using the texture-based technique, and characters extending beyond the bounding boxes are filled in. However, the authors don't offer any quantitative analysis regarding the performance enhancement when using this hybrid approach.

Antani et al. [10, 17] proposed a multi-pronged approach for Text Information Extractions and its all processes. Author(s) used different techniques to reduce the chance of failure. The results of each detection and localization technique may be a set of bounding boxes surrounding the text regions. The results from the individual techniques are then united to provide the final bounding boxes. Author(s) utilize algorithms by Gargi et al. [13,14], Chaddha et al. [29], Schaar-Mitrea and De With [30], and LeBourgeois [31]. Chaddha et al. presented use of texture energy to classify 8×8 pixel blocks into text or non-text blocks. The sum of absolute values of a set of DCT coefficients from MPEG data is thresholded to catgorised a block as text or non-text. Schaar-Mitrea and P. De With's formula [30] was originally developed for the classification of graphics and video. Antani et al. [10] modified this to classify 4×4 blocks as text or non-text. The number of pixels of pixels in block with the similar color is compared with threshold values to classify the block.

LeBourgeois [31] used a threshold value based on the sum of the gradients of every pixel's neighbors. Antani's effort to utilize and merge many traditional approaches would appear to provide an inexpensive performance for general image documents. Later, in his thesis, Antani [10] gave additional thought to the merging strategy. Gandhi et al. [32] presented planar motion model for scene text localization. The motion model parameters are then used to compute the structure and motion parameters. Perspective correction is performed using the calculable plane normal, leading to a far better OCR performance. However, since only motion data is employed for text segmentation, the algorithm requires more processing time. The employment of additional information, like texture or tonal data, might enhance the results.

Niblack's [33] adaptive thresholding method is used to separate non-text regions. Author(s) investigated the relationship between the OCR accuracy and and the resolution and found that the best OCR results were obtained when using a factor of 4 in image interpolation. Li and Doermann [34] also used a multiple frame-based text image enhancement technique, wherever consecutive text blocks are correctly tracked registered using a pure translational motion model. The text blocks are correctly tracked, the mean square errors of two consecutive text blocks and motion trail information are used. A sub-pixel accuracy registration is additionally performed bi-linear interpolation to reduce the noise sensitivity of the two low-resolution text blocks. However, this could reduce non-text regions only if text and background have different movements. Lastly, a super-resolution-based text enhancement scheme is additionally given for de-blurring scene text, onto convex sets (POCS)-based method [35].

## 2.3 Tracking, Extraction and Enhancement

This section discusses tracking, extraction, and enhancement Techniques. Due to many reasons, utility, enhancement speedup, etc., tracking of text in video has not been studied in great extends. Locating text in images, such as low resolution and complex backgrounds, these topics need to be more investigation.

To enhance the system performance, it's necessary to consider temporal changes in a frame sequence. The text tracking stage will serve to verify the text localization results. Additionally, if text tracking may be performed in a shorter time than text detection and localization, this might speed up the general system. In cases wherever text is occluded in different frames, text tracking can be facilitate recover the original image.

Lienhart [21,22] represented a block-matching algorithm, that is an international standard for video compression like H.261 and MPEG, and used temporal text motion info to refine extracted text regions. The minimum mean absolute difference is used as the matching criterion. Each localized block is checked on whether or not its fill factor is higher than a given threshold value. For every block that meets the specified fill factor, a block-matching algorithm is performed.

Antani et al. [17] and Gargi et al. [14] utilize motion vectors in an compressed video MPEC-1 bit stream for trackingthe text, based on the strategies of Nakajama et al. [62] and Pilu [63]. This methodology is implemented on the P and that I frames in MPEG-1 video streams. The original bounding box is then moved by the total of the motion vectors of all the macroblocks that correspond to current bounding box. Li et al. [34] presented a approach or many circumstances, including scrolling, captions, text printed on an athlete's jersey, etc. Author(s) used the sum of the square difference for a pure translational motion model, based on multi-resolution matching, to reduce the computational complexity. The text contours are used to stabilize for additional complex motions for tracking process. Edge maps are generated using the canny operator for larger text block. Once a horizontal smearing method to cluster the text blocks, the new text position is extracted. However, since a translational model is used, this methodology isn't appropriate to handle scale, rotation and perspective variations.

Sawaki et al. [38] projected a techniques for adaptively acquiring templates of degraded characters in scene images involving the automated creation of 'context-based image templates' from text line images. Zhou et al. [39,40,41] use their own OCR algorithm based on surface fitting classifier and n-tuple classifier. we address the issue of text extraction enhancement for generating the input to an OCR algorithm.

Text enhancement techniques are often divided into two categories: single frame-based or multiple frame-based. Several thresholding techniques are developed for still images. However, these strategies don't work well for video sequences. Based on the fact that text typically spans several frames, varied approaches that use a tracking operation in consecutive frames are proposed to enhance text quality. Such enhancement strategies are often effective when the background movement is completely different from the text movement.

Sato et al. [42] used a linear interpolation technique to magnify small text at a better resolution for commercial OCR software package. This methodology is unable to clean the background when both the text and background moving at the same time. After the image enhancement stages, four specialized character-extraction filters are applied based on correlation and a recognition-based segmentation methodology is used for character segmentation.

Li et al. [43,44] given many approaches for text enhancement. Author(s) use the Shannon interpolation technique to enhance the image resolution of video frames. The image resolution is increased an extension of the Nyquist sampling theorem and it's determined whether or not the text is normal or inverse by comparing it with a global threshold and background color.

Antani et al. [10] used number of algorithms for extracting text. Author(s) developed a binarization algorithm similar to Messelodi and Modena's approach [24]. To refine the binarization result, a filtering stage is utilized based on color, size, spacial location, topography, and shape. These extraction algorithms operate on both the original image and its inverse image to detect text regions. Chen et al. [45] proposed a text enhancement methodology that uses a multi-hypotheses approach.

## 3. TEXT BASED IMAGE VIDEO INDEXING AND RETRIEVAL

The unique properties of video collections (e.g., multiple sources, noisy features and temporal relations) examine the performance of these retrieval methods in such a multimodal environment, and identify the relative importance of the underlying retrieval Components. Li and Doermann [47], presents text-based video indexing and retrieval by expanding the semantics of a query and using the glimpse matching method to perform approximate matching. Cees et al.[49], identify three strategies to select a relevant detector from thesaurus. Jawahar, et. al. [51] presents search approach that based on the textual information present in the video. Textual information regions are identified within the frames of the video. Based on textual content present in the images, video is annotated. OCRs are used to extract the text from the video. Author(s) proposed an approach that provides matching at the image-level and avoided an OCR. Based on query string matching videos are retrieve from database and sort it based on relevance.

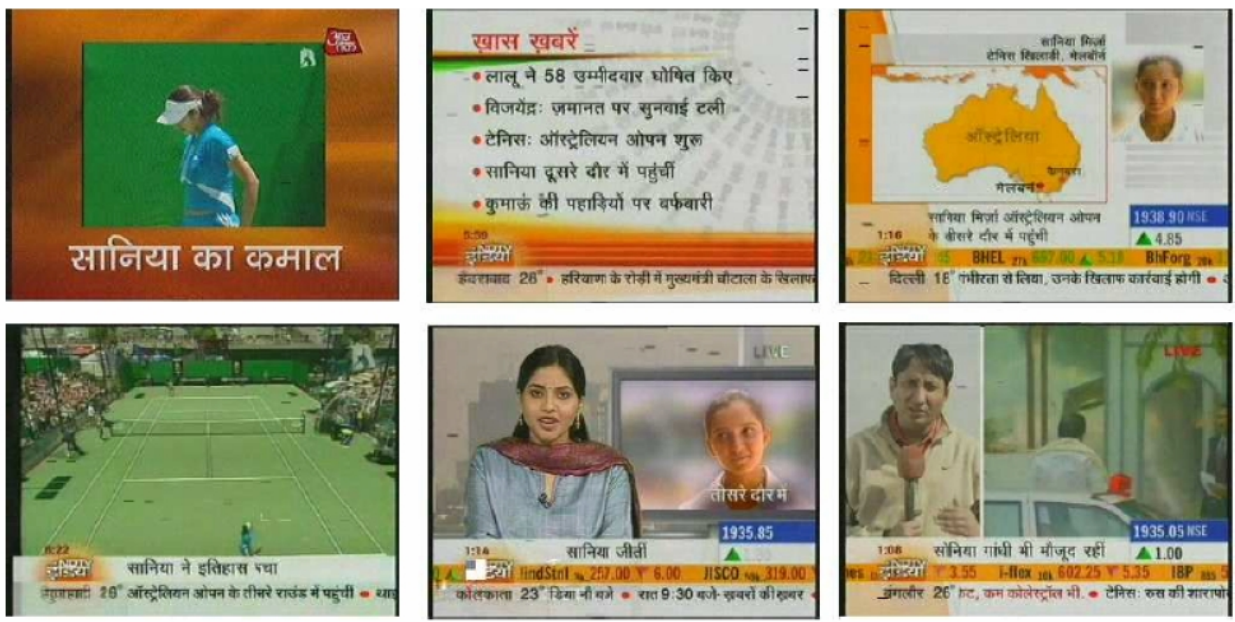

Fig 5. Result for Query video Sania in devnagari script  (Courtesy Jawahar, C. V, et. al. [51])

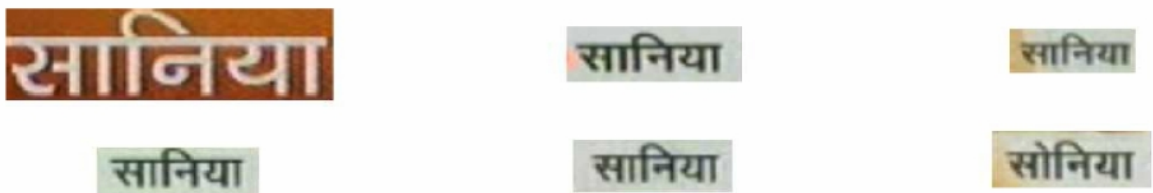

Fig 6. Query word "Sania" in devnagari script  from Fig 5. frames (Courtesy Jawahar, C. V, et. al. [51])

## 4. CONCLUSION

In this paper we are provided the comprehensive literature review of text extraction in images and video as well as text based image and video retrieval. In the literature, a large number of algorithms have been presented but didn't find any techniques which provide satisfactory performance. The different information Sources (e.g., color, texture, motion, shape, geometry, etc).  are used for text. By merging the different sources of information we and enhance the performance of a text extraction system and text based video retrieval systems.